\documentstyle[11pt,aaspp4,epsf]{article}
\begin{document}


\newcommand{\msun}{\mbox{${\cal M}_\odot$}}
\newcommand{\lsun}{\mbox{${\cal L}_\odot$}}
\newcommand{\kms}{\mbox{km s$^{-1}$}}
\newcommand{\HI}{\mbox{H\,{\sc i}}}
\newcommand{\HeI}{\mbox{He\,{\sc i}}}
\newcommand{\mhi}{\mbox{${\cal M}_{HI}$}}
\def\hst{{\it HST}}
\def\rsun{{\rm\,R_\odot}}
\def\etal{{\it et al.}}
\newcommand{\OIII}{\mbox{O\,{\sc iii}}}
\newcommand{\HII}{\mbox{H\,{\sc ii}}}
\newcommand{\NII}{\mbox{N\,{\sc ii}}}
\newcommand{\SII}{\mbox{S\,{\sc ii}}}
\newcommand{\nan}{Nan\c{c}ay}
\newcommand{\skp}{\mbox{  }\\}
\newcommand{\bmv}{\mbox{B--V}}
\newcommand{\bmr}{\mbox{B--R}}
\newcommand{\darkV}{$\frac{{\cal M}_{HI}}{L_{V}}$}
\newcommand{\dark}{$\frac{{\cal M}_{HI}}{L_{B}}$}
\newcommand{\am}[2]{$#1'\,\hspace{-1.7mm}.\hspace{.0mm}#2$}
\newcommand{\as}[2]{$#1''\,\hspace{-1.7mm}.\hspace{.0mm}#2$}

\title{The Extraordinary `Superthin' Spiral Galaxy UGC~7321. I. Disk
Color Gradients and
Global Properties from Multiwavelength Observations}
\vskip0.5cm
\author{L. D. Matthews\altaffilmark{1,}\altaffilmark{4,}\altaffilmark{5}}

\author{J. S. Gallagher, III\altaffilmark{2,}\altaffilmark{4,}\altaffilmark{5}}

\author{W. van Driel\altaffilmark{3,}\altaffilmark{4,}\altaffilmark{5}}

\altaffiltext{1}{National Radio Astronomy Observatory, 520 Edgemont Road, 
Charlottesville, VA 22903, 
Electronic mail: lmatthew@nrao.edu}
\altaffiltext{2}{University of Wisconsin, Department of Astronomy, 
475 N. Charter St.,
Madison, WI 53706-1582}
\altaffiltext{3}{Unit\'e Scientifique Nan\c{c}ay, CNRS USR B704,
Observatoire de Paris, 92195 Meudon, France}
\altaffiltext{4}{Guest observer at the WIYN Observatory, which  is a 
joint facility of the University of
Wisconsin-Madison, Indiana University, Yale University, and the
National Optical Astronomy Observatories.}
\altaffiltext{5}{Visiting Astronomer, Kitt Peak National
Observatories, which is operated by the Association of Universities
for Research in Astronomy, Inc. (AURA) under a cooperative agreement
with the National Science Foundation.}

\singlespace
\tighten
\begin{abstract}
We present $B$- and $R$-band 
optical imaging and photometry, H$\alpha$ narrow-band imaging,
near-infrared $H$-band imaging, 
and \HI\ 21-cm spectroscopy of
the nearby ($V_{h}$=407~km/s), 
Sd spiral galaxy UGC~7321.  UGC~7321 exhibits
 a remarkably thin stellar disk with no discernible bulge
component. The galaxy has a very diffuse, low surface brightness disk,
which appears to suffer relatively little
internal extinction in spite of its nearly edge-on geometry 
($i\approx$88$^{\circ}$). If seen face-on, UGC~7321 would have an observed
central $B$-band surface brightness of only
$\sim$23.4~mag~arcsec$^{-2}$. 

The UGC~7321 disk shows significant $B-R$ color gradients in both the
radial and vertical directions:
$\Delta(B-R)\ge$0.80 magnitudes along the galaxy major axis, and
$\Delta(B-R)$ as large as 
0.45 magnitudes is observed parallel to 
the galaxy minor axis. These color gradients cannot be explained solely by
dust and are indicative of changes in
the mix of stellar ages and/or metallicity 
as a function of both radius and height
above the galaxy plane. The outer regions of the UGC~7321 disk are too
blue to be explained by low metallicity alone ($B-R\le0.6$), and must be
relatively 
young. However, the galaxy also contains stellar populations
with $B-R>$1.1, indicating it is not a young or recently-formed galaxy. 

The disk of UGC~7321 is not a simple exponential, but exhibits a light
excess at small radii, as well as distinct surface brightness zones.
Despite its organized disk structure, 
many of the  global properties of UGC~7321
($M_{B}=-17.0$; \mhi=1.1$\times10^{9}$\msun; \dark=1.1 \msun/\lsun;
$W_{20}$=233~\kms; $h_{r}$=2.1~kpc) 
are reminiscent of a dwarf galaxy. Together the 
properties of UGC~7321
imply that it is an under-evolved galaxy in both a dynamical and
in a star-formation sense.

\end{abstract}


\section{Background}
Ogorodnikov (1958) and 
Vorontsov-Vel'yaminov (1967) were among 
the first to draw attention to a subset of
edge-on spiral galaxies with  disk axial
ratios as extreme as 30:1\footnote{These disk axial ratios were
measured on blue photographic survey plates; however, we find
 the axial ratios tend to 
become less extreme when  measured at a fixed isophote on 
deeper, red-sensitive CCD images. For this reason, measurements of 
vertical disk scale heights offer
 a more robust means of quantifying
the thickness of edge-on galaxies.}   and having little or no
discernable bulge
component. Goad \& Roberts (1981) dubbed these galaxies ``superthins'' 
and recognized
that spiral galaxies selected on the basis of their thin stellar disks and 
extreme axial ratios also tend to
possess a variety of other intriguing properties. 

For example, Goad \& Roberts (1981) discovered that superthins have low
emission-line intensity ratios ([\NII]/H$\alpha$ and [\NII]/[\SII]),
which in a modern perspective suggests that superthins are moderate
ionization, low metallicity galaxies more like irregulars than spirals
in terms of their \HII\ region properties (see also Bergvall \&
R\"onnback 1995). Similar trends were confirmed for larger samples by
Karachentsev \& Xu (1991) and Karachentsev (1991).

In spite of having ``classic'' double-horned global \HI\ profiles
(e.g., Matthews \& van Driel 1999), often superthins exhibit
slowly-rising, solid-body rotation curves throughout most or all of
their stellar disks (Goad \& Roberts 1981; Karachentsev \& Xu
1991; Karachentsev 1991; Cox {\it et al.} 1996; Makarov \etal\ 1997,1998;
Abe {\it et al.} 1999). 
Such rotation curves are generally characteristic of
late-type dwarf and irregular 
galaxies rather than normal spirals (cf. Casertano \&
van Gorkom 1991).

Not only are superthins fascinating objects in their own right, but 
the very nature of these galaxies makes them extremely valuable objects
for exploring a variety of fundamental astrophysical problems,  ranging from 
disk formation and evolution scenarios, to the interpretation of
high-redshift galaxy populations, to constraining the stability
and dark matter contents of disks (e.g., Karachentsev 1989; Zasov
\etal\ 1991; Dalcanton \& Schectman 1996). In spite of this,
to date only a handful of 
detailed studies of individual superthins have
been undertaken (e.g., Gallagher \& Hudson 1976; 
Bergvall \& R\"onnback 1995; Cox \etal\ 1996;
Matthews 1998; Abe {\it et al.} 1999).

Measurements of the vertical structure and
disk color gradients are of particular interest in the case of superthin 
spirals, as these can provide important
clues to their evolutionary history.
Vertical structure measurements provide vital constraints on disk
stability, dynamical heating mechanisms, and dark matter contents
(Matthews 1998; Matthews 1999, hereafter Paper~II), 
and when combined with measures of global properties
and disk color gradients, information on the star-formation
histories and dynamical evolutionary histories of the galaxies can be gleaned.

UGC~7321 is a superthin galaxy ideally suited to such investigations. Since
it is nearby (D$\sim$10~Mpc; Matthews {\it et al.} 1999b), 
its structure can be reasonably resolved from the ground.
In addition, UGC~7321 appears to suffer little from 
internal extinction (see below), and it is nearly
exactly edge-on ($i\approx$88$^{\circ}$), hence projection effects
will have a minimal influence on derived vertical structural parameters 
(e.g., de Grijs, Peletier, \& van der Kruit 1997). Finally, the analysis
of UGC~7321 is simplified compared with other edge-on spirals
due to the absence of a bulge component. 

Here we present
measurements of the global  properties of UGC~7321 based on
new multiwavelength observations
including $B$- and $R$-band imaging and photometry, narrow-band H$\alpha$
imaging, near-infrared $H$-band imaging, and \HI\ 21-cm pencil beam
spectroscopy.  We also examine \HI\ aperture synthesis data 
obtained by Rots, Roberts, \& Goad (private communication).
We use our datasets to measure the radial and vertical
color gradients in the
disk of UGC~7321 (Sect.~3) and briefly discuss what the color
gradients reveal about the 
evolutionary history 
of this galaxy. 
In a second paper (Paper~II), we present
measurements of the vertical scale height of the UGC~7321 disk. There we
combine vertical structural measurements
with  results from the present
work in order to further
constrain the stability and dynamical 
history of UGC~7321.

\section{Imaging Data}
\subsection{Near-Infrared Imaging}

We obtained near-infrared (NIR) $H$-band imaging observations of
UGC~7321 in May 1997 using IRIM on the 2.1-m telescope at the Kitt Peak
National Observatory\footnote{Kitt Peak National Observatory
is operated by the Association of Universities for Research
in Astronomy, Inc. under
contract with the National Science Foundation.}. IRIM employs
a 256$\times$256 HgCdTe NICMOS3
detector with
\as{1}{09} pixels, yielding a field-of-view of  \am{4}{6} per side.
The detector
has a gain of 10.5~$e^{-}$/ADU and a readnoise of  $\sim$37~$e^{-}$ per
pixel rms. The observations were taken in the presence of 
moderate cirrus, hence
our $H$-band data cannot be reliably photometrically calibrated.

We obtained sets of 9 co-added, 20-$s$ exposures
of UGC~7321 at each of 35 telescope pointings. To
insure accurate flatfielding and sky subtraction, we shifted the
galaxy to a different position on the detector before each series of
exposures.
In addition, interspersed throughout the target observations,
we obtained several series of ``blank''
sky  exposures at regions  adjacent to the
galaxy. Linearization corrections to the data
were made
automatically during data acquisition.

Data reduction was accomplished using standard IRAF\footnote{IRAF is
distributed by the National Optical Astronomy Observatories, which is
operated by the Associated Universities for Research in Astronomy,
Inc. under cooperative agreement with the National Science Foundation.} tasks.
To remove the sky background, a
median  was made of
several groups of temporally consecutive object and blank
sky exposures. From these, a   median-averaged  combination
 of 25 dark frames was subtracted  to remove the structure
present in the detector dark current. Finally, the dark-corrected
sky frame was
subtracted from all dark-corrected object frames.

Flatfielding was accomplished by dividing by a night sky flat composed
of the median of all dark-corrected sky and object exposures. Lastly,
the flatfielded object exposures were aligned and co-added.
Our resulting $H$-band image of UGC~7321 is shown in Fig.~1. Note the
striking thinness of the stellar disk and the complete absence of 
a bulge component even in the NIR. At the disk center, the vertical
exponential scale
height is only $\sim$140~pc (Paper~II)! 

Fig.~2 shows
the $H$-band data in the form of a contour map. Because the disk is so
thin, the contours have been
stretched in the $z$-direction for clarity. Note that the contours are
quite regular, even near the galaxy center, suggesting that
there is not  significant dust absorption in the $H$-band.

In Fig.~3 we plot a major axis profile extracted from the $H$-band
data. From this profile, we see that the disk 
brightness of UGC~7321 
falls off somewhat more steeply on the east side (the righthand side
of Fig.~3) than on the west side. It is unlikely this can be
attributed to internal extinction effects (which should be small 
at NIR wavelengths), or to
flatfield errors (which are quite small in our $H$-band images), hence this
suggests a slight lopsidedness in the stellar disk of UGC~7321.

Arrows overplotted on Fig.~3 denote ``zones'' between which the slope of the
brightness profile change fairly abruptly.  
These zones can also be seem from visual inspection of
both our NIR and the optical images and are an indication that the
disk surface brightness of UGC~7321 deviates somewhat from a smooth,
exponential distribution (see also Sect.~2.2.3). This makes it
difficult to characterize the $H$-band major axis profile in terms of
a single exponential scale length. 
As shown by Matthews \& Gallagher (1997),
this phenomenon is common among extreme late-type spiral galaxies,
although the effect is generally lost if the light profile is
azimuthally averaged. Matthews \& Gallagher (1997) suggested these
surface brightness zones 
may be a signature of limited dynamical evolution in some of the
smallest and faintest spirals
(see also Sect.~2.3.3). 

\subsection{Optical CCD Imaging and Photometry}

\subsubsection{Data Acquisition}
We obtained photometrically calibrated $B$- and $R$-band
CCD imaging observations of UGC~7321 in June 1997 using the 3.5-m
WIYN\footnote{The WIYN Observatory is a joint facility of the
University of
Wisconsin-Madison, Indiana University, Yale University, and the
National Optical Astronomy Observatories.} telescope
at Kitt Peak. In addition, we obtained
narrow-band H$\alpha$ and uncalibrated $B$-and $R$-band observations of the
galaxy with the WIYN telescope in April 1997.

The direct imaging camera on the WIYN telescope employs a thinned STIS
2048$\times$2048 CCD with \as{0}{2} pixels, yielding a field-of-view of
$\sim$7$'$ per side. The CCD gain was 2.8~$e^{-}$ per ADU, and
readnoise was $\sim$8~$e^{-}$
per pixel rms. Exposure times were 900$s$ in $B$, 600$s$ in $R$, 
and 1500$s$ in H$\alpha$. Seeing during both the April and
June observations
was $\sim$\as{0}{6} FWHM. The WIYN telescope takes advantage of the
best seeing on Kitt Peak through a combination of thermal controls,
active optics, and dome ventilation systems. This,
combined with the  small pixel scale of the WIYN CCD
imaging camera, makes WIYN an ideal facility
for obtaining the data necessary for detailed studies of disk 
structure and color gradients of galaxies.   

Our CCD images were reduced using standard
IRAF tasks. Individual frames were overscan corrected
and bias subtracted in the usual manner. Flatfielding was accomplished
by dividing by a mean of 5 dome flats taken in the appropriate
filter. The WIYN
CCD contains a bad column which was corrected in our images by
interpolation from the two adjacent columns.

Both the April and the June observations were obtained during dark time.
Photometric calibration  for the June run was accomplished by
monitoring two standard star fields from Landolt (1992) at two
different air masses and then applying standard color term corrections
to convert from WIYN natural magnitude system to the Landolt
system. The total formal dispersion in our WIYN magnitude zero points
is $\pm$0.03 magnitudes; the validity of our transformations
was  checked via comparisons with
other recent WIYN photometric solutions. Errors in the colors are
approximately $\pm$0.02 magnitudes. Sky brightnesses during our
observations were $\mu_{B,sky}\sim$22.1~mag~arcsec$^{-2}$ and 
$\mu_{R,sky}\sim$19.8~mag~arcsec$^{-2}$.

Our $R$-band image of UGC~7321 is shown in Fig.~4 \& 5. Note the very
diffuse, low
surface brightness (LSB) appearance of the disk in spite of its edge-on
geometry. Several background
galaxies are clearly visible through the disk (e.g., Fig.~6), implying
internal extinction is low.

\subsubsection{H$\alpha$ Emission}
Our continuum-subtracted H$\alpha$ image of UGC~7321 is shown in
Fig.~7.  Due to cirrus, we were unable
to obtain a flux calibration for this frame. 
However, qualitatively, this image shows that H$\alpha$ emission is
present out to radii of at least $r=\pm$\am{2}{5} (where $r$ is the
distance from the galaxy center measured along the disk major axis in
the plane of the sky).

We see in Fig.~7 that the bulk of the 
H$\alpha$ emission in UGC~7321 is
confined to a region near the disk midplane roughly 4$''$ thick,
although diffuse emission is clearly visible at $z$-heights of up to
$\pm8''$ at galactocentric radii $|r|\le$\am{1}{0}.
We note however that there is a relative
dearth of emission within a region offset 8$''$ 
to the east of the disk center (Fig.~8).
This ``gap'' is $\sim 8''$ wide and contains faint plumes
of emission extending
to $z\ge\pm$7$''$ out of the galaxy plane. We also see marginal
evidence of several other 
thin filaments which are more extended in $z$-height.
Deeper narrow-band 
images are clearly needed to establish whether the central
H$\alpha$ feature could be a
signature of blowout from the disk, and whether UGC~7321 may indeed contain a
diffuse, ionized medium extended to large $z$-heights, analogous 
to that seen in
more luminous edge-on spirals (cf. Dettmar 1995 and references therein).

\subsubsection{Surface Photometry}
We performed surface photometry on UGC~7321 using routines from
the IRAF STSDAS analysis package. With the ``ellipse'' program,
we fit a series
of 9 concentric ellipses to the UGC~7321 images. Position angle
and ellipticity of the ellipses were determined from the outermost
galaxy isophotes and were kept fixed throughout the
fitting. The galaxy center was chosen to be the position of  peak
brightness; this location was identical in the $B$ and $R$ (and $H$)
frames.
Foreground stars, cosmic rays, and background galaxies 
within the aperture were removed via background interpolation 
using the IRAF task
``imedit''. Sky values were determined by measuring the sky counts
in each of several rectangular apertures at various locations on the image
(see Matthews \& Gallagher 1997).

Because UGC~7321 is so close to edge-on, elliptical apertures are not
ideal fits to the galaxy isophotes. Nonetheless, these apertures allow us
to measure  aperture magnitudes and colors for UGC~7321, and
permit an estimate of the mean exponential scale length of its
disk.  Although the
ellipticity of the fitted isophotes is an unreliable method of
determining the
inclination of near edge-on galaxies, our derived value
($i\approx88^{\circ}$)
is consistent with the inclination we
estimate from 
the slight asymmetry of the dust features along the galaxy major axis.

Our derived photometric parameters for UGC~7321
are given in Table~1. Errors for the  aperture magnitude 
measurements were computed
following Matthews \& Gallagher (1997) and take into account sky,
flatfield and Poisson errors, as well as the scatter 
in the photometric solution. Large-scale 
flatfield errors are the dominant source
of uncertainty; the maximum amplitude of the flatfield variations was
$\sim$1\% of sky in both $B$ and $R$.  
Our aperture magnitudes are in good
agreement with previous photoelectric values reported by Tully (1988;
$m_{B}$=13.86) and de Vaucouleurs \& Longo (1988; $m_{R}$=12.99).

Fig.~9 shows our
azimuthally averaged $B$-band radial surface brightness profile of 
UGC~7321. Although
we noted in Sect.~2.1 that the disk of UGC~7321 is slightly lopsided
and shows deviations from a perfect exponential, an
azimuthally-averaged brightness profile permits determination of a
``mean'' scale length which is useful for offering a global
characterization of the size scale of the disk.

Because UGC~7321 is viewed close to edge-on, one must take into account
projection effects in analyzing the projected light profile and in
deriving an exponential scale length.
For a disk with an exponential radial brightness distribution viewed
exactly edge-on ($i=90^{\circ}$), 
the projected radial brightness profile
along the galaxy midplane ($z$=0) is
expressed as 
\begin{equation}
L(r)=L_{0}(r/h_{r})K_{1}(r/h_{r})
\end{equation}
\noindent where $K_{1}$ is the
modified Bessel function of first order 
(e.g., van der Kruit \& Searle 1981). At
small radii ($r/h_{r}<<1$), this expression can be approximated as
\begin{equation}
L(r)\approx L_{0}[1 + (r^{2}/2h_{r}^{2}){\rm ln}(r/2h_{r})].
\end{equation}
\noindent Note this implies  a slight
flattening of the light profile will be observed at small radii
compared with a simple, unprojected $e^{-r/h_{r}}$ function.
At large radii ($r/h_{r}>>1$), one can write
\begin{equation}
L(r)\approx
L_{0}(\pi r/2h_{r})^{1/2}{\rm exp}(-r/h_{r})
\end{equation}
\noindent (see van der Kruit \& Searle 1981). Because of the
$\sqrt{r}$ term, this produces a slightly less steep light profile
than a pure exponential function with the same scale length. As a result, a
scale length derived from fitting a simple, unprojected 
exponential profile to an edge-on disk will be 
overestimated. At $r=2$-$3h_{r}$ this effect is $\sim$10\% (e.g., van
der Kruit \& Searle 1981).

To derive the scale length for UGC~7321, we have fitted a model
projected exponential light profile to our data. Our
model is overplotted on Fig.~9. For the purpose of deriving a scale length,
the slight deviation of UGC~7321 from the edge-on ($i=90^{\circ}$) case is not
significant. From our fit we derive a scale length of
$h_{r,B}$=44$''\pm2$ 
(compared with $h_{r,B}$=51$''\pm5''$ we derive
from simply fitting the function
$L(r)=L_{0}e^{-r/h_{r}}$ to the disk at intermediate $r$).

A comparison between the data and our model (Fig.~9) shows that in spite of
lacking a bulge component, UGC~7321 exhibits
a light excess at small radii compared with the prediction
of an exponential disk. 
The $B$-band {\it measured} central surface brightness of UGC~7321
(before correction to a face-on value) is 21.6~mag~arcsec$^{-2}$,
while that predicted from extrapolation of the exponential fit to
small $r$ is $\sim$0.35 magnitudes fainter.  Interestingly, this
excess is even more pronounced in the $R$-band and 
appears to correspond to a distinct region in the color map of
UGC~7321 (see Sect.~3).
This strengthens the suggestion that 
UGC~7321 appears to have a
multi-component disk that is more complex than a simple,  pure
exponential (see also Sect.~2.1).
In addition, we note that 
at $r>120''$ our observed light profile falls off faster than the
projected exponential 
model, suggesting the stellar disk of UGC~7321 
may be truncated (cf. Barteldrees
\& Dettmar 1994). However, this latter trend should be confirmed with deeper
observations on a wide-field CCD where sky subtraction and
flatfielding can be accomplished with greater accuracy.

\subsubsection{Discussion: The LSB Nature of UGC~7321}
For an optically thin 
disk which is exponential in both the $r$ and $z$ directions,
and which is observed inclined at 90$^{\circ}$,  
the face-on central surface brightness will be
$h_{z}/h_{r}$ times the observed edge-on value. Here  $h_{z}$ is the
exponential scale height (cf., van der Kruit \& Searle
1981). Transformations from observed surface brightness values to face-on
values at other radii may be derived from Equations 1-3
above. 
Using a
thin-disk approximation (see the Appendix) we have derived the additional
corrections
to these values required for
the case where a is disk observed at an inclination slightly less than
90$^{\circ}$. 
After also taking into account internal
extinction corrections (see Sect.~3.2.2), we then find that if
projected to face-on, the disk of UGC~7321 would have an {\it observed} 
$B$-band central surface brightness of $\sim$23.4~mag~arcsec$^{-2}$, 
an extrapolated central surface
brightness of $\sim$23.8~mag~arcsec$^{-2}$, and a mean total disk surface
brightness $\bar\mu_{B}\sim$27.6~mag~arcsec$^{-2}$.  
Thus UGC~7321 is a very LSB galaxy, and much of its disk 
would likely be nearly invisible if viewed
closer to face-on. As shown below, the internal extinction in UGC~7321
appears to be quite low, thus its LSB appearance must
result from a rather low current star formation rate.

Further
evidence that the seemingly ``anemic'' nature of UGC~7321 is due to
minimal current star
formation comes from comparing its blue luminosity to its
far-infrared 
luminosity. Using the {\it IRAS} 60$\mu$m and 100$\mu$m fluxes for
UGC~7321 from
the NED database and Sage (1993), respectively, 
and following the prescription of
Rice \etal\ (1988), we derive $L_{FIR}$=7.8$\times10^{7}$\lsun, a
value nearly two orders of magnitude fainter than the mean for Scd/Sd
spirals in the UGC catalogue (Roberts \& Haynes 1994). Using our blue
luminosity (corrected for internal extinction), we find that 
$L_{FIR}/L_{B}\approx$0.08 for UGC~7321. A comparison with Rice \etal\ (1988)
reveals that such low $L_{FIR}/L_{B}$ ratios are
commonly found only in two classes of galaxies: (1) 
old, red early-type spirals, dE's and S0's with few young stars; (2) 
gas-rich, extreme late-type spirals with 
diffuse, extremely LSB disks. 
Examples of the latter group of objects include the
nearby galaxies NGC~4395 (Sd~IV), NGC~45 (Sdm~IV), and IC~2574
(Sm~IV-V). All three of these galaxies have $L_{FIR}/L_{B}\le$0.10,
\dark$>$1, blue optical colors, and highly transparent disks (e.g.,
Sandage 1961; Matthews \etal\ 1999a).

\section{Color Maps}
Using our $B$- and $R$-band WIYN data, we have produced a $B-R$ color
map of UGC~7321 (Fig.~10). A variety of structure is evident in this
map, including significant 
$B-R$ color gradients in both the vertical and radial
directions. We discuss these features in detail below.

The global $B-R$ color of UGC~7321 is 0.99, as 
measured within the outermost observed
isophote 
(25.2~mag~arcsec$^{-2}$ in $B$ before correction for inclination),
and after correction  for Galactic foreground extinction. 
This $B-R$ color
is a typical value for a normal, late-type spiral (cf.
Lauberts \& Valentijn 1988; de Blok, van der Hulst, \& Bothun 1995).
This suggests that in spite of its edge-on geometry, the disk of
UGC~7321 does not suffer severely
from internal reddening due to dust. We return to 
this issue below.

\subsection{The Nuclear Region of UGC~7321}

Near the
center of the UGC~7321 disk, our color map reveals  a small, very red
($B-R\approx$1.5)
nuclear feature, only a few arcseconds across (Fig.~10). This feature
is offset $\sim$5$''$
to the east of the disk center as determined from the brightness peak
of the $R$- and $B$-band images. One possibility is 
that we are seeing the signature of an embedded nucleus. However,
although compact star cluster nuclei are common in late-type
low-luminosity spirals with diffuse, LSB disks (Matthews \& Gallagher 1997;
Matthews \etal\ 1999a),
we do not see any direct evidence of such a feature in UGC~7321
in either our WIYN images
or in images obtained with the {\it Hubble Space Telescope}
(Matthews \etal\ 1999b). 
Nonetheless, 
the location of the compact red feature in the
$B-R$ color map is near to that of 
the peculiar H$\alpha$ emission features shown in Fig.~8.
The optical longslit spectrum of UGC~7321 obtained by
Goad \& Roberts (1981) also shows a possible kinematic disturbance in 
the H$\alpha$
emission at this location. These are hints 
that  some interesting physical processes
are at work near the center of the UGC~7321 disk. Further
investigation of 
this region  via high resolution
spectroscopy, deep H$\alpha$ imaging, and NIR colors 
may be fruitful.

Surrounding the compact red feature, our color maps reveal a more
extended red region ($B-R\sim$1.2), visible to $r\approx\pm20''$ on
either side of the disk center and showing a rather abrupt
boundary (see also below). Intriguingly, the extent of this region 
corresponds very
closely to the region over which we observe a 
light excess over a pure exponential disk fit
(Sect.~2.2.3). This raises the possibility that this red central region
might possibly represent an
ancient central starburst,  the core
of the original protogalaxy, or perhaps even a 
kinematically distinct disk subsystem analogous to
the bulge of normal spirals.

\subsection{Radial Color Gradients}

\subsubsection{General Trends}

Cutting into the red central region discussed above are thin blue bands of
stars visible along the  midplane
of the galaxy. These bands grow both thicker and bluer with increasing
distance from the galaxy center.
At $|r|=$20$''$, this layer has $B-R\approx$1.05, 
reaching $B-R\approx$0.85 at $|r|$=\am{1}{0}, 
$B-R\approx$0.80 at $|r|=$\am{1}{5},
$B-R\approx$0.45$\pm0.10$ at $r$=\am{2}{7}, and $B-R\approx$0.55$\pm$0.10
at $r=-$\am{2}{7}. Thus, from the nuclear region (where $B-R\approx
1.5$) to the outer disk edge we
see a total
$B-R$ color change of up to 1.05 magnitudes along the major
axis of UGC~7321. This is illustrated in Fig.~11$a-d$,
which shows $B-R$ color
profiles of UGC~7321 extracted along the major axis, as well as
\as{1}{6} north (through the red nuclear feature),
\as{1}{2} south, and \as{4}{6} south  
relative to 
the major axis, respectively. The extracted profiles were averaged
over  12-pixel-wide
strips and then smoothed, hence the red color of the compact red
central disk feature
is slightly subdued in the profiles shown in Fig.~11. Because  we are
interested in accurately measuring colors in the faintest regions of the outer
disk, we extracted our color profiles from our
our uncalibrated $B$ and
$R$ images, which have slightly superior flatfields 
to our photometrically-calibrated  data; maximum amplitude 
variations of the flatfields are
$\sim$1\% of the sky in
$B$ and $\sim$0.5\% of the sky in $R$. We then  used comparisons
between high
signal-to-noise inner disk regions
to calibrate the absolute colors from  the
photometrically-calibrated images. The maximum 
uncertainties expected from the
combination of sky subtraction and flatfield errors are overplotted as
dotted lines on Fig.~11.
Although signal-to-noise in the outer disk is low, the profiles
extracted at all four positions parallel to the major axis
show regions with $B-R\le$0.60, suggesting these blue outer disk
colors are real.

Finally, we draw attention to the presence of a faint, thicker, but highly 
flattened disk
of unresolved stars visible
in our color map surrounding the UGC~7321 disk at
$|r|\le$\am{2}{0}. This component has $B-R\approx$1.1 and 
shows little change in
color with galactocentric distance (Fig.~12). Based upon the
observations of Galactic globular clusters (see Secker 1995), old,
metal-poor stellar populations are expected to have $B-R>$0.8, hence
the highest $z$-height stars in UGC~7321 appear to represent an ``old disk''
population (see also below).

\subsubsection{Gauging the Effects of Dust}

Bluing as a
function of increasing galactocentric distance is a well-known feature of
spiral galaxy disks (e.g., de Jong 1996). De Jong
(1996) has  demonstrated through Monte Carlo simulations
that in face-on spirals, this observed radial bluing  generally
cannot be accounted for by dust. He shows that for
realistic models, dust creates color gradients  of less than 0.3
magnitudes in $B-R$, and argues  instead
that the observed radial color changes in spirals
are indicative of star formation progressing radially outwards
in disks over time, leading to stellar age and metallicity gradients. 
This picture is consistent with a number of 
semi-analytic galaxy formation models where galaxy disks are built ``from the
inside out'' (e.g., White \& Frenk 1991;
Mo \etal\ 1998). Nonetheless, the radial color gradient we observe in
UGC~7321 ($\Delta B-R\sim1$) is significantly larger than typical
gradients in the de Jong sample (where typically $\Delta B-R\sim$0.6
magnitudes, even when bulge light is included). Moreover, in most
spirals observed
edge-on, radial color gradients are generally  found to be small
(e.g., Sasaki 1987; Aoki \etal\ 1991; Wainscoat, Freeman, \& Hyland 
1989; Bergvall
\& R\"onnback 1995) or negligible (e.g., Hamabe \etal\ 1980; Jensen \&
Thuan 1982; de Grijs 1998).

Unfortunately, in edge-on spiral galaxies, 
color gradients become more difficult to interpret physically,
particularly since internal
extinction can be quite significant near the galactic plane. The
net effect  can be both an alteration in the galaxy luminosity profile,
and a reddening of the observed optical 
colors which can vary as a function of $r$ and $z$. Therefore in order
to determine what fraction of the color gradients
we observe in UGC~7321  are due to true
age and/or metallicity
gradients, we must include an assessment 
of  the role of dust in this galaxy.

In
practice,  accurately quantifying the effects of dust on the observed
colors and luminosity distribution in a galaxy is a complicated
problem (e.g., de Jong 1996; 
Xilouris \etal\ 1999). 
Moreover, standard assumptions about the nature and
distribution of dust in normal giant spirals are unlikely to be applicable to 
low-metallicity LSB galaxies  (cf. Han 1992).  As a result, the effects
of internal extinction in a given galaxy are ideally derived 
using a combination of sophisticated
radiative transfer models and empirical measurements (e.g., Xilouris
\etal\ 1999).  
Such modelling beyond the scope of the
present work,  so here we derive some fundamental constraints on the
effects of dust in UGC~7321 using a very simple model.

Although some authors have argued that the internal extinction in LSB
galaxies may be almost negligible (e.g., McGaugh 1994;
Tully \etal\ 1998), these
claims have only rarely been tested in edge-on systems (e.g., Goad \&
Roberts 1981; Kodaira \&
Yamashita 1996; Bergvall \& R\"onnback 1995; Karachentsev 1999). 
We have already mentioned several lines of evidence
that the internal extinction in UGC~7321 is low (regularity of the 
$H$-band
isophotes; visibility of background galaxies through the disk; low
$L_{FIR}/L_{B}$ ratio; correspondance between the galaxy center in $H$-
and $B$-band). Nonetheless, visible inspection of our WIYN images shows
the galaxy is clearly not completely devoid of dust, and a number of individual
clumps (possibly molecule-rich dark clouds)
can be seen in our optical images (e.g., Fig.~5; see also Matthews
\etal\ 1999b). In spite of this, 
the fact that we resolve many of these individual dust clumps instead of
seeing a uniform dust lane 
implies that  the disk of UGC~7321 is not optically thick.

Another test of the optical thickness of the UGC~7321 disk comes from
a comparison between
its optical rotation curve (derived from longslit spectroscopic
measurements of H$\alpha$ emission from \HII\ regions), with a
rotation curve derived 
from \HI\ measurements. From optical longslit measurements,
Goad \& Roberts (1981) found UGC~7321 
to have a slowly rising, solid-body rotation curve throughout much of
its stellar disk. If a galaxy disk is optically thick,
the
rotation curve may appear solid body regardless of its intrinsic shape,
as an artifact of one's inability to observe \HII\ regions at small
galactocentric radii
 (Goad \& Roberts 1981; Byun 1993). However, a rotation curve of
UGC~7321 derived from \HI\ aperture synthesis measurements (Sect.~4.2)
confirms that the slow rise of the rotation curve throughout
the stellar disk 
is indeed intrinsic (see Fig.~19, discussed below). 
Once again we conclude the disk of UGC~7321 is not optically thick.

Dust affects the color and luminosity profiles of galaxies through
both the scattering and absorption of optical photons. 
However, the maximum color change (i.e., maximum reddening) occurs
in the pure extinction case. In addition, Byun, Freeman, \& Kylafis 
(1994) have
shown that scattering effects become decreasingly important with
increasing inclination (see also Bianchi, Ferrara, \& Giovanardi
1996). 
Since here we are primarily
interested in estimating the effects of dust on the color
gradient, and since we are
in the optically thin regime, we can simplify the problem by
considering  a  model of extinction due to a foreground dust
screen (see Disney, Davies, \& Phillips 1989). 
For a given amount of dust, the Screen model produces
more reddening than  a more realistic model of stars mixed with
dust, hence it establishes an upper limit to $E(B-R)$ as a
function of radius.

To evaluate $E(B-R)$, we begin by attempting to attribute 
as much of our observed color
gradient as possible to dust.
We further assume that, to first order,
dust extinction in the $H$-band is negligible (e.g., Fig.~2). In the
case of zero intrinsic radial color gradient,
a dust-free version of UGC~7321 would
exhibit no radial color gradient in $R-H$, hence we can use the
changes we do observe in the $R-H$ color to estimate the amount of
extinction at a given radius.
We do not attempt to reproduce the  clumpy nature of the dust in our
model (which again, would decrease any reddening effects), 
but assume the dust distribution is roughly exponential in
the radial and vertical directions (e.g., Wainscoat, Freeman, \&
Hyland 1989; Xilouris \etal\ 1999). In this case the
extinction along an interval from $r$ to $\delta r$ can be
expressed as
\begin{equation}
\delta A_{\lambda,i}(r,z)=\lbrace\begin{array}{r}
A_{\lambda,0}e^{(-r/h_{r,d}-|z|/h_{z,d})}\delta r~~~~~~~~~~r\le R_{max} \\ 
0~~~~~~~~~~~~~~~~~~~~~~~~~~~~~~~~r>R_{max} 
\end{array} 
\end{equation}
\noindent where $A_{\lambda,0}$ is the absorption in magnitudes per
unit length at a
given wavelength, $r=R_{max}$ corresponds to the edge of the
stellar disk, $h_{r,d}$ is the scale length of the dust, and $h_{z,d}$
is the scale height of the dust (see also Wainscoat, Hyland, \&
Freeman 1989).
 
Fig.~13 shows a plot of the ``pseudo'' $R-H$ color profile of UGC~7321
 along the major axis interval $r$=0--120$''$. Here the $R$-band data were
smoothed to match the resolution of the $H$-band observations.
Because our $H$-band data are not photometrically-calibrated, 
we have chosen an arbitrary
normalization such  that our ``pseudo'' $R-H$ color is 0 near
$r$=80$''$.  In Fig.~13, we see that near $r$=0, 
the $R$-band light is depressed by roughly 40\%
relative to the $H$-band, corresponding to an $R$-band extinction
$A_{R,0}\approx$0.55 magnitudes per kiloparsec. 
Assuming the extinction law of Bouchet
\etal\ (1985), we derive
$A_{B,0}$=0.89 magnitudes per kiloparsec. 

From Fig.~13 we  estimate a 
scale length of the dust distribution of $h_{r,d}\approx$40$''$, which
appears consistent with the observed dust distribution in our images. 
Although in giant spirals, it is
often assumed the dust has a similar scale length to the old stellar
disk (e.g., Kylafis \& Bahcall 1987), or even extends beyond it (e.g.,
Xilouris \etal\ 1999),
this does not appear to hold in UGC~7321. We see an abrupt end to the
presence of resolved dust clumps beyond radii $|r|\approx 80''$ in both
our WIYN images, and in {\it Hubble Space Telescope} images obtained
by Matthews
\etal\ (1999b). Moreover, background galaxies are readily visible through the
UGC~7321 disk beyond these radii (Matthews \etal\ 1999b), implying
extinction has dropped appreciably. For the vertical scale height of
the dust, we adopt $h_{z,d}$=\as{2}{3}, which is half the scale height
of the old stellar disk (Sect.~3.3; see also Paper~II).

Using Equation 4, we can now derive an
extinction-corrected $B-R$ 
color profile for UGC~7321 along its major axis (Fig~14$a$) and offset
\as{1}{6} to the north (Fig~14$b$).  We see that even  the extinction-corrected
radial color gradient remains pronounced ($\Delta B-R\sim0.80$ mag). This  
demonstrates that {\it a significant fraction of the observed
radial color gradient in UGC~7321 appears to be due to stellar population
and/or metallicity gradients.}

\subsubsection{Discussion and Interpretation}

The presence of the observed radial disk 
color gradients in UGC~7321 has several
implications. Regions with $B-R<0.8$ are too blue to be accounted for
by old, metal-poor stellar populations (e.g., Secker 1995), but
instead must result from a mixture of ages. Without resolving these
populations, we cannot uniquely
constrain the mix of stars present. However, the bluest
colors we reliably measure in UGC~7321 ($B-R<0.6$) are comparable, for
example,  to
young regions in the moderate metallicity giant Magellanic galaxy
NGC~4449 (Bothun 1986) and to the predicted colors of the  
younger outer regions of a smoothly evolving galaxy with a high halo spin
parameter and an age of $\sim$10~Gyr, as in the models of 
Jimenez \etal\ (1998). In either case, it is likely that youthful stars are a
significant contributor to the blue colors of the outer disk of
UGC~7321 (see also Bell \etal\ 1999). This assertion is strengthened
by the presence of H$\alpha$ emission in the outer disk (Fig.~7) and
the existence of resolved supergiant stars at these radii (Matthews
\etal\ 1999b).

We note that edge-on, superthin LSB galaxies like UGC~7321
are especially valuable for studies of the outer regions of disks, as
they may contain some of the most pristine, unevolved galaxian disk
environments observable in the nearby universe. However, because these disk  
regions are so optically thin, they 
would become very difficult to detect in galaxies viewed at
lower inclinations. For example, if viewed near face-on,
the portions of the  UGC~7321 disk where $B-R\le$0.6  would
have an observed $B$-band surface brightnesses of only 
$\ge$27.6~mag~arcsec$^{-2}$.  Such faintness levels are only rarely
reliably achieved in  galaxy imaging surveys 
due to combinations of short integration times,
scattered light, limited detector fields-of-view, and flatfield
uncertainties (cf. Morrison, Boronson, \& Harding 1994; Lequeux \etal\ 1996). 

In spite of the potentially very young outer disk 
regions of UGC~7321, its moderately red central disk (where
$B-R\approx$1.2) and its somewhat thicker ``old disk'' 
(where $B-R\approx$1.1) both indicate that
UGC~7321 is likely to contain 
stellar populations with ages in excess of 10~Gyr
(see Jimenez \etal\ 1998). As demonstrated above, accounting for 
a reasonable amount of reddening due to dust does
not affect this conclusion. The red stellar components in 
UGC~7321 imply that this is not a young 
galaxy 
presently undergoing its first epoch of star formation. 

Our observed colors and color gradients in UGC~7321 are similar to
those found in other LSB spirals by de Blok, van der Hulst, \& Bothun
(1995; see also Impey \& Bothun 1997). A consistent model for UGC~7321
is then that of a disk galaxy which has made stars slowly and is now
seen to be under-evolved relative to typical giant spirals (see also
O'Neil \etal\ 1997; Jimenez
\etal\ 1998). This would account for the low gas metallicity of
UGC~7321 (Goad \&
Roberts 1981),
the large \HI\ mass fraction (Sect.~4), and the low density of stars
(i.e., the
low optical surface brightness disk).
Thus while LSB disks like UGC~7321 may be {\it
unevolved} galaxies, we concur with the suggestion of Jimenez {\it
et al.} (1998) that {\it LSB galaxies are not
necessarily young.}

In order to preserve its relatively 
strong radial color gradients over many Gyr, it appears that
viscous evolution in the disk of UGC~7321 has been minimal.  In giant
spirals, viscous evolution (which redistributes angular
momentum in the disk) is often argued to be responsible for the 
relatively smooth exponential luminosity profiles of most disk galaxies
(e.g., Lin \& Pringle 1987). This process is also expected to
partially smooth color gradients due to stellar age and metallicity 
changes with 
radius (e.g., Firmani, Hern\'andez, \& Gallagher 1996). 
However, such a mechanism may not be able to
work efficiently in galaxies like UGC~7321, due to both the low disk surface
density (which considerably extends dynamical timescales) and the 
slowly rising rotation curve [which produces limited shear in the
inner disk regions
(Lin \& Pringle 1987; see also Matthews \& Gallagher 1997)]. The 
gradients in the $H$-band major-axis profile of the UGC~7321 disk 
(Sect.~2.1) and
the deviation of the azimuthally-averaged profile from a pure
exponential disk at small $r$ (Sect.~2.2.3) may be
additional
signs that  viscous evolution has been limited in this galaxy compared
with normal spirals.

Finally, we note that at intermediate galactic radii, 
the eastern side of UGC~7321 has a slightly bluer mean color than the
western
side.
This result is insensitive to reasonable sky subtraction and
flatfield errors. One explanation could be that we 
are seeing the projection of  a spiral arm(s), although it remains
uncertain how much spiral structure a thin
disk like UGC~7321 might actually sustain, since spiral structure is
believed to be an efficient disk heating mechanism (e.g., Lacey 1991
and references therein). Moreover, at least some self-gravity
is needed for spiral arm formation, but if the disk of UGC~7321 were
completely self-gravitating, it would be highly unstable to ``firehose''
instabilities and hence to 
vertical thickening (Zasov, Makarov, \& Mikailova 1991; Matthews 1998).
An alternative is that
there are  asymmetries in the stellar distribution of UGC~7321
due to patchy star-formation
(see also Gerritsen \& de Blok 1999). 

\subsection{Vertical Color Gradients}
The presence of vertical color gradients in galaxies offers
an important clue toward the dynamical evolution of galaxian disks, since such
gradients are predicted to occur as a consequence of dynamical heating
processes  (e.g., Just,
Fuchs, \& Wielen
1996). Unfortunately, vertical color gradients are very difficult to
measure in most edge-on spirals due to dust, contamination from
the bulge component, and the effects of atmospheric 
seeing. To date, only a few
analyses of edge-on spirals have uncovered non-negligible vertical
color gradients (e.g., Wainscoat, Freeman, \& Hyland 1989; Bergvall \&
R\"onnback 1995), while many workers  have reported such gradients to
be very small or negligible  (e.g., Jensen \& Thuan 1982; de Grijs,
Peletier, \& van der Kruit 1997).

Fig.~15$a$-$h$ illustrates  the vertical color profiles of the UGC~7321 disk
at various galactocentric
distances.
At $r=0$, we see
very little $z$ color gradient: $B-R\sim$1.2 at a range of $z$ values
(Fig.~15$a$), although a slight asymmetry
is visible.
At $r=6''$ (Fig~15$b$), we  see the addition of a  
peak with $B-R\sim$1.45. Note its displacement from the disk
midplane; this peak is in 
the vicinity of the red nuclear feature discussed in Sect.~3.1. 
At $r$=10$''$ (Fig~15$c$), the vertical color gradient
is again fairly flat ($B-R\sim$1.2), with  a very slight bluing visible
near $z$=\as{1}{5}.
At $r=$\am{0}{5} (Fig~15$d$),
the high $z$ regions of the disk have nearly the same
color as those at $r$=0 (this is due to the ``old disk'' 
of red stars
discussed above), but at
small $z$ values, the disk has become much bluer than at
smaller galactocentric radii.
Finally, with further increasing radius, while the color at high $z$ continues
to stay nearly constant, the color at small $z$ grows increasingly
bluer (Fig.~15$e$-$h$),
consistent with the radial bluing observed along the major axis of the
galaxy in Fig.~11.

Vertical color gradients of the type we observe over much of the disk
of  UGC~7321 (i.e. redder with
increasing $z$ height)
are predicted to occur due to dynamical
disk evolution in which older stellar populations acquire higher velocity
dispersions over time due to heating processes  (e.g., Fuchs \& Wielen 1987;
Just, Fuchs, \& Wielen 1996).
UGC~7321 represents one of the
few examples  of this
type of vertical color structure being directly 
observed in a galaxy disk. Thus even the most dynamically cold
examples of nearby galaxy disks appear to have undergone some dynamical
heating. Nonetheless, the exact mechanism by
which older stars are dynamically heated and how they are 
redistributed as a function time
remain uncertain (see review by Lacey 1991) and different mechanisms
may operate in different galaxies, depending upon factors like
environment, degree of self-gravity, and the size and number of
molecular clouds.
The relative simplicity of UGC~7321, and
its status as a relatively
unevolved galaxy make this an ideal system to help
place important constraints on these issues. We explore this 
further in Paper~II.

Using the extinction parameters derived in Sect.~3.2.2, 
we can also estimate the effects of dust reddening on our observed
vertical color gradients. We adopt for a scale height of the dust half
the scale height of the old stellar disk (Xilouris \etal\ 1999). 
Using the stellar scale height of the old disk
derived in Paper~II, this yields $h_{z,d}$=\as{2}{3}.
Fig.~16  shows an  example of one of our vertical color profiles
corrected for dust reddening. We see that at the galactocentric radii
where the vertical color gradients become the strongest, correction
for dust reddening acts  to slightly increase $\Delta(B-R)$ along
the vertical direction.

\section{HI Observations}
\subsection{HI Pencil Beam Mapping}
In June 1997 we used
the \nan\ Decimetric Radio Telescope to obtain a
7-point pencil-beam map of UGC~7321 in the  21-cm line of neutral
hydrogen (\HI). At the declination of
UGC~7321, the \nan\ Radio Telescope has a FWHM beam size of
approximately 4$'$E-W$\times$23$'$N-S. Other
details regarding the  \nan\
telescope may be found in, e.g., Matthews, van Driel, \& Gallagher
(1998).

UGC~7321 was observed  at 7 different telescope pointings, including a
position corresponding to the optical center of the galaxy, and
positions offset 2$'$, 4$'$, and 6$'$ east and west of center, respectively.
The observations
were obtained in total
power (position-switching) mode using consecutive pairs
of two-minute on- and two-minute
off-source integrations. Total integration times were 2-3 hours
at each pointing. For these observations, the
autocorrelator was divided into two pairs
of cross-polarized receiver banks, each with 512 channels and a 6.4~MHz
bandpass. This yielded a channel spacing of 2.6~\kms, for an effective
velocity resolution of $\sim$3.3~\kms.

Fortuitously, UGC~7321 is oriented along nearly an E-W line, so the
\nan\ telescope provides sufficient spatial resolution along
this direction to obtain
crude 
information about the \HI\ distribution and kinematics of the
galaxy.
The individual spectra obtained at each telescope pointing are shown in
Fig.~17 and
the  resulting global spectrum in Fig.~18. 
Note the global profile appears quite symmetric. Table~1 
gives our derived global
\HI\ parameters. Errors were computed following Matthews, van Driel,
\& Gallagher (1998). 
Our values are in good agreement
with the recent single-dish \HI\ observations of UGC~7321 
obtained by Haynes \etal\ (1998)
using the Green Bank 140-ft telescope.

Our \nan\ observations show that
the \HI\ gas in UGC~7321 clearly extends beyond the
optical galaxy, confirming the results  of Hewitt, Haynes, \& Giovanelli
(1983). Significant flux is detected at the pointings both 4$'$E
and 4$'$W of the galaxy, giving a lower limit for the for \HI-to-optical
diameter ratio of $D_{HI}/D_{opt}\ge$1.25. In addition, we detect a
small amount of \HI\ flux
($\sim$1.1~Jy~\kms) in our observation 6$'$W of the galaxy
center. From the \nan\ data alone, it
is difficult to assess whether this flux is due to a sidelobe
contamination or real extended emission, 
since the strength of the \nan\ telescope sidelobes varies
significantly with the
hour angle of the source (Guibert 1973).

\subsection{HI Aperture Synthesis Data}
After our \nan\ observations were obtained, we had the opportunity to
examine \HI\ aperture synthesis data
of  UGC~7321 obtained by Rots, Roberts, \& Goad
(private communication) using the Very Large Array (VLA).  These 
observations were obtained in C array in 1981 using 18 antennas,
hence they do not achieve the same sensitivity limits as 
modern VLA data, but they still offer a useful complement to 
our \nan\
dataset. Thirty-one independent velocity channels were used for the
observations, with a velocity resolution of $\sim$10.3~\kms\ per
channel. The
FWHM
beamwidth was $\sim$12$''$. 

Using the VLA data, we derived a
position-velocity diagram for  UGC~7321. By assuming  that at
each point along the major axis the maximum velocity traces the
rotation, we have derived the rotation curve shown in Fig.~19. 
This figure confirms
the slowly-rising, solid body nature of the rotation of UGC~7321
throughout
its stellar disk, as was first seen in the optical rotation curve of
Goad \& Roberts (1981). Although beam-smearing may be
expected to slightly decrease the amplitude of our derived \HI\ rotational
velocities, this effect is not expected to exceed a few kilometers per second,
or to be significant beyond 2--2.5 beam diameters from the disk center
(e.g., Swaters 1999), 
hence it cannot explain the shallow shape of the inferred rotation curve.
Finally, we note that Fig.~19 suggests that the rotation curve
of UGC~7321 does not begin to flatten until near the edge of the
stellar disk. 
This is consistent
with UGC~7321 being a galaxy whose dynamics are 
dominated by a dark halo even at small galactocentric radii 
(e.g., Matthews 1998). However,
deeper, more sensitive \HI\ aperture synthesis measurements would be
valuable for further constraining the detailed shape of the outer
rotation curve.

The integrated \HI\ map of UGC~7321 derived by Rots, Roberts, \& Goad 
(private communication)
confirms a slight extension of the \HI\ on the west side of
the galaxy compared with the east side, consistent with our \nan\
data.
It also permitted us to
measure an \HI\ diameter of $D_{HI}\sim$\am{7}{1} for UGC~7321 at the
limiting  \HI\ column density  of the observations ($N_{HI}\sim
7\times10^{20}$~atoms~cm$^{-2}$). Thus
$D_{HI}/D_{opt}\ge$1.2, in agreement with our \nan\ lower limit.

\subsection{Discussion}
It is somewhat difficult to accurately
compare the \HI\ extent of UGC~7321 with other
spirals since the commonly-quoted
$D_{HI}/D_{25}$ ratio (where $D_{25}$ is measured at a
face-on-corrected isophote) in not particularly meaningful for a galaxy like
UGC~7321, whose face-on central surface brightness is only
$\sim 23.4$~mag~arcsec$^{-2}$. Moreover, we cannot unambiguously
translate our limiting observed \HI\ column density into an \HI\
surface density due to projection effects, thus it is difficult to
accurately measure $D_{HI}$ at some canonical surface density (e.g.,
1~\msun~pc$^{-2}$). 

In terms of optical scale length,
the \HI\ in UGC~7321 extends to at least 5.5$h_{r}$.  
Although this value may increase  with more sensitive VLA
observations, such 
an extent is still quite normal for both high and
low surface brightness late-type spirals
having similar rotational velocities to UGC~7321 
(see Fig.~8 of de Blok, McGaugh, \& van der Hulst 1996).

If we assume both the stellar and the \HI\ disks of UGC~7321 are
optically thin, and hence both similarly enhanced in surface
brightness due to their edge-on projection, then we see that the
maximum \HI\ extent of the disk relative to the maximum optical extent
is typical for Sd spirals (e.g., Hewitt, Haynes, \& Giovanelli
1983).  We can more directly compare with two other Sd superthins
previously mapped
in \HI:
$D_{HI}/D_{opt}\sim$1.2 was also found by Cox \etal\ (1996) for the
superthin spiral UGC~7170 and by Abe \etal\ (1999) for the superthin
IC~5249, both at similar limiting \HI\ column densities and optical
surface brightnesses. 

Ignoring internal extinction effects, 
the \dark\ ratios of UGC~7321, UGC~7170, and IC5249 are
2.6, 3.4, and 2.3 (in solar units), respectively, and the superthin
spiral ESO~146-014 studied by Bergvall \& R\"onnback (1995) has
\dark$\sim$2.8. If we assume all have internal extinctions similar to
UGC~7321 ($A_{B,i}\sim0.89$ mags; Sect.~3.2.2), the implied \dark\
ratios are still significantly
larger than typical values for normal Sd spirals derived by Roberts \&
Haynes (1994; \dark$\sim$0.63),
although not as extreme as some of the unevolved dwarf galaxies
studied by van Zee \etal\ (1995)
or some of the extreme late-type LSB spirals in the sample of  Matthews
\& Gallagher (1997) which have \darkV\ as high as 10 
(see also Salzer \etal\ 1991; 
Matthews, van Driel, \& Gallagher 1998).

Van der Hulst \etal\ (1993) suggested that low gas surface densities
are responsible for the low star-formation efficiencies in LSB
galaxies (see also de Blok, McGaugh, \& van der Hulst 1996; Gerritsen
\& de Blok 1999). Although we cannot unambiguously recover the radial 
\HI\ surface
density distribution in UGC~7321 due to its edge-on geometry
(cf. Olling 1996), we can compute a mean \HI\ surface density within
the stellar disk. Adopting the definition $\overline{\Sigma}_{HI}=$\mhi$/\pi
R^{2}_{opt}$, where $R_{opt}$ is the linear optical diameter of the disk
(e.g., Roberts \& Haynes 1994) we find
$\overline{\Sigma}_{HI}$=5.3~\msun~pc$^{-2}$. This value is considerably
lower than the median of this quantity found by Roberts \& Haynes
(1994) for Scd/Sd spirals in the
UGC catalogue: $\overline{\Sigma}_{HI}$=9.80~\msun~pc$^{-2}$, but is
consistent with the 
$\overline{\Sigma}_{HI}$ ratios of several of 
the LSB galaxies in the sample of
de Blok, McGaugh,
\& van der Hulst (1996).

\section{Summary}
We have presented $B$- and $R$-band imaging and photometry,
near-infrared $H$-band imaging, narrow-band H$\alpha$ imaging, and
\HI\ 21-cm line measurements of the nearby, edge-on  Sd spiral galaxy
UGC~7321.

UGC~7321 is a ``superthin'' galaxy, with an extremely
dynamically cold stellar disk and no discernible bulge component, even
in the near-infrared. In spite of its
edge-on orientation, UGC~7321 is visibly quite diffuse and it is clear
that its intrinsic optical surface brightness is 
quite low. The dust content of UGC~7321 also appears to 
be small, hence we argue that the ``anemic'' appearance of the galaxy
results from
the  low-level of current star formation rather than severe
internal extinction.

UGC~7321 exhibits significant  $B-R$ color gradients in the radial
direction:
measured from the disk center to its edges
edge, $\Delta (B-R)\sim$1.05 magnitudes along the galaxy major
axis. Dust alone cannot explain the large gradient. Using a simple
extinction model we find
$\Delta (B-R)_{cor}\sim$0.80 after correction for internal reddening.
This is somewhat larger
than the color gradients typically observed in normal giant spirals
(cf. de Jong 1996), and suggests significant stellar population
gradients in the disk of UGC~7321.

The outermost disk regions of UGC~7321 have $B-R\le$0.6, suggesting
they are composed of stellar populations that include a significant
fraction of young stars. 
However, UGC~7321 also contains a
population of old stars with $B-R\ge$1.1, indicating it is not a
young or recently-formed  galaxy. The rather strong radial
segregation of these populations suggests that the galaxy has evolved
quite slowly and perhaps that viscous
evolution has not operated efficiently in this system.

UGC~7321 also exhibits appreciable vertical color gradients:
$\Delta B-R$ as large as 0.45 magnitudes was measured parallel to the
minor axis. This is a
reflection of a concentration of young blue stars along the galaxy
midplane, and a population of the older, red stars at larger scale heights. 
This type of age segregation is predicted to occur due to
dynamical heating processes in spiral disks, 
but UGC~7321 represents one of the few examples of it being directly observed
in an external galaxy. This  implies that even
dynamically cold disks like UGC~7321 have undergone some dynamical heating.

The stellar light distribution of the UGC~7321 disk cannot be characterized by
a single exponential function. A light excess over the prediction of an
exponential model is seen at small radii, and in both optical and NIR
wavelengths, the disk appears to
contain distinct surface brightness ``zones'' between which the slope
of the brightness profile changes. This may be an additional signature
of minimal viscous evolution. It also brings into question the
prediction of some semi-analytic galaxy models (e.g., Dalcanton,
Spergel, \& Summers 1997) that exponential stellar disks are a natural
product of the disk formation process.

In spite of its regular, organized disk,
many of the global properties of UGC~7321, including its luminosity,
\HI\ content, rotational velocity, and \dark\ ratio 
(Table~1), as well as its slowly-rising, solid-body 
rotation curve  (Fig.~19)
are more typical
of a dwarf irregular galaxy than a normal Sd spiral. The origin of such
vastly different disk morphologies in an otherwise similar
physical parameter space remains unclear, but it may place an
important constraint on galaxy formation and evolution models.

Together the properties of UGC~7321 suggest that it is an under-evolved
galaxy in both a dynamical and in a star-formation sense. Nonetheless,
this galaxy clearly demonstrates that
even seemingly simple ``pure disk'' galaxies like the superthins are
highly complex systems. 

\acknowledgements
{We are grateful Arnold Rots and Mort Roberts 
for providing us with the fully-calibrated
VLA \HI\ data of UGC~7321. We also thank
Mike Merrill for assistance with the IRIM observations
at Kitt Peak and Wanda Ashman for creating the artwork for the Appendix. 
This research was partially funded by the Wide Field and 
Planetary Camera 2 (WFPC2) Investigation Definition Team,  which is
supported at the Jet Propulsion Laboratory (JPL) via the National
Aeronautics and Space Administration (NASA) under contract No.
NAS7-1260.  The \nan\ Radio Observatory is the department {\it Unit\'e
Scientifique \nan} of the {\it Observatoire de Paris} and is associated with 
the French {\it Centre National de Recherche  Scientifique} (CNRS) as 
the {\it Unit\'e de Service et de Recherche} (USR), No. B704. The \nan\
Observatory also gratefully acknowledges the financial support of the 
{\it R\'egion Centre} in France.  This research made use of the
NASA/IPAC Extragalactic Database (NED), which is operated by the Jet
Propulsion Laboratory, California Institute of Technology, under
contract with the National Aeronautics and Space Administration.}

\appendix
\section{Estimating the Effects of
Inclination on the Observed Central Surface Brightness of a Disk}
When deriving central surface brightness 
measurements of galaxies, it is desirable to convert measured values
to face-on values in order that galaxies observed at different
inclinations can be  meaningfully
compared. 
However, for highly-inclined galaxies, the
approximation that $\mu_{face-on}=\mu_{observed}+2.5{\rm log}(a/b)$
(e.g., Freeman 1970) breaks down, and alternative corrections must be used.
Here we derive a simple formula for this purpose, using a ``thin
disk'' approximation.

In our model we assume the galaxy disk is optically thin and 
has an exponential brightness
profile along the $r$ direction of the form $L(r)=L_{0}e^{-r/h_{r}}$,
where $L_{0}$ is the luminosity volume density and $h_{r}$ is the disk
scale length. We also assume the disk has a finite but
non-negligible thickness of 2$h_{z}$, and a constant brightness along the
$z$ direction (see Fig.~20). For such a disk, the central surface
brightness if viewed edge-on is expressed as 
\begin{equation}
L(0)_{i=90}=L_{0}\int _{-\infty}^{+\infty}e^{-r/h_{r}}dr=2L_{0}h_{r};
\end{equation}
\noindent if viewed face-on, the central surface brightness
would be simply $L(0)_{i=0}=2L_{0}h_{z}$.

Now suppose our model disk is viewed at some arbitrary angle $i'<90^{\circ}$
through a line of sight along
a vector {\bf x} (see Fig.~21). We define $i$ as the angle between the
line of sight and a normal to the disk. Now the integrated brightness of the
disk along a path from $x=0$ to $x=x_{max}$ becomes
\begin{equation}
L(0)_{i=i'}=2L_{0}\int_{0}^{x_{max}}e^{x{\rm sin}i/h_{r}}{\rm sin}i~dx.
\end{equation}
\noindent Taking the integral, and making the substitution
\begin{equation}
x_{max}=h_{z}/{\rm cos}i,
\end{equation}
\noindent we find that
\begin{equation}
L(0)_{i}=2L_{0}h_{r}[1 - e^{-h_{z}{\rm tan}i/h_{r}}].
\end{equation}
\noindent To apply this approximation to a real galaxy, one can substitute the
measured radial and vertical scale heights for $h_{r}$ and $h_{z}$
respectively, provided that the conditions $h_{z}<<h_{r}$ and
\begin{equation}
(90 - i') > {\rm tan}^{-1}\left(h_{z}/x_{max}\right)
\end{equation}
\noindent both hold.

\newpage

\figcaption{$H$-band image of UGC~7321 obtained with IRIM on the KPNO
2.1-m telescope. Field size is $\sim$\am{4}{6}$\times$\am{1}{0}.
East is on the right
of the image, north on top. A number of 
foreground stars and cosmic rays have been
cleaned from the image. Note the complete
absence of a bulge component, even in the near-infrared.}

\figcaption{Contour map produced from the $H$-band image of
UGC~7321. The image has  been stretched in the $z$ direction for
illustration purposes. The contours are spaced in units of 20
counts. Note the regularity of the contours, even near the galaxy
center, suggesting dust absorption is minimal in the $H$-band.} 

\figcaption{$H$-band major axis profile of UGC~7321, averaged over a
3-pixel-wide strip. The axes are
distance along the galaxy major axis, in arcseconds, 
versus the logarithm of the
intensity, in arbitrary units. The arrows mark the locations of
changes in slope of the brightness profile.}

\figcaption{$R$-band image of UGC~7321 obtained with the WIYN
telescope. Image size is $\sim$\am{6}{5}$\times$\am{1}{0}. 
East is on the right, north
on top. Seeing was $\sim$\as{0}{6} FWHM.}

\figcaption{Close-up of a \am{1}{8} long portion of the $R$-band WIYN
image of UGC~7321. The optical center of the  galaxy corresponds to
the large black patch on the righthand side of the image. White areas
show dust clumps in the disk. Note the absence of a true dust
lane. }

\figcaption{Close-up of the western edge of the UGC~7321 disk in the $R$
band, showing a background galaxy  visible though the disk, at
approximately \am{2}{5} west of the disk center.}

\figcaption{Continuum-subtracted H$\alpha$ image of UGC~7321 obtained
with the WIYN telescope. Scale and orientation are as in Fig.~4.}

\figcaption{Detail of the H$\alpha$ emission structure centered at
$\sim 8''$ east of the galaxy center. This image section
is roughly 50$''$ long.
Note the gap in the emission at the center of this frame
and the faint plumes extending  out of the disk midplane.}

\figcaption{Azimuthally-averaged $B$-band surface brightness profile
of UGC~7321 obtained from fitting a series of 9 concentric ellipses. 
Axes are distance from the galaxy center in arcseconds,
versus average $B$-band surface brightness in magnitudes per square
arcsecond. The dashed line represents the best exponential fit to the
data. No inclination or extinction corrections have been applied.}

\figcaption{$B-R$ color map of the inner \am{3}{8} of 
UGC~7321 obtained from our WIYN
data. Seeing in both the $B$ and $R$ data was $\sim$\as{0}{6}.
Black (near the nuclear region) corresponds to $B-R\approx1.5$; 
dark grey (seen  surrounding the nuclear region): $B-R\approx$1.2; 
mottled black and white (seen in the thicker disk surrounding the
galaxy): $B-R\approx$1.1; white (as seen along the disk major axis and
in the outer disk regions): $B-R<$0.85.}

\figcaption{$(a)$ $B-R$ color
along the major axis of UGC~7321, averaged over a 12-pixel-wide strip. The
data have been
smoothed by a factor of 15 along the $r$-direction. The
axes are distance from the galaxy center in
arcseconds versus $B-R$ color in magnitudes, corrected
for Galactic foreground extinction. $(b)$ Same as $(a)$, but offset
\as{1}{6} north of the major axis, and running through the very red
nuclear feature (see Sect.~3.1). $(c)$ Same as $(a)$, but offset
\as{1}{2} south
from the major axis. $(d)$ Same as $(a)$, but offset \as{4}{6} south
from the major axis. The very red feature near $r=-135''$ is due to a
background galaxy seen through the disk.}

\figcaption{Same as Fig.~11$(a)$, except the 
profile was extracted  
10$''$ north of the major axis. This cut runs through the
thicker ``old disk'' of stars visible in Fig.~10. Note there is
very little color gradient as a function of $r$. Beyond $r=\pm$75$''$ our
extracted 
profile no longer traces the ``old disk'' due to the curvature of the
galaxy isophotes.}

\figcaption{``Pseudo'' $R-H$ color along the major axis of UGC~7321, in
the interval $r\le 120''$. The $R$ data were smoothed to the $H$-band
resolution, and the $R$ and $H$ intensities have been
arbitrarily normalized such that $R-H\approx$0 near $r$=80$''$.}

\figcaption{$(a)$ Same as Fig.~11$(a)$, but with internal reddening
correction applied. $(b)$ Same as Fig.~11$(b)$, but with internal
reddening correction applied. }

\figcaption{$B-R$ vertical color profiles extracted parallel to the
minor axis of UGC~7321. The profiles are averaged over 15-pixel-wide
strips and smoothed along the $z$ direction.
Axes are distance from the galaxy midplane, in
arcseconds, versus $B-R$ color, corrected for Galactic foreground
extinction. $(a)$ Minor axis profile ($r$=0); $(b)$ 
$r$=5$''$ east; $(c)$ $r$=10$''$ east; $(d)$ $r$=\am{0}{5} east; $(e)$
$r$=\am{1}{0} east; $(f)$ $r$=\am{1}{25} east; $(g)$ $r$=\am{1}{5}
east; $(h)$ $r$=\am{2}{0} east. }

\figcaption{Vertical $B-R$ color profile at $r$=\am{1}{25} east of the
disk center, 
corrected for internal reddening. The uncorrected profile is
overplotted as a dashed line.}

\figcaption{\HI\ pencil beam map along the major axis of  
UGC~7321 obtained with the \nan\
Radio Telescope. The center
panel shows the spectrum obtained at the
optical center of the galaxy, and the remaining panels show spectra
obtained at  pointings offset west and east of center, respectively,
at multiples of one-half beamwidth.  
Axes are heliocentric radial velocity, in
\kms$\times10^{-2}$, versus flux density in millijanskys.}

\figcaption{The global \HI\ profile of UGC~7321. 
 Data from the seven telescope pointings shown in Fig.~17 were
combined following Matthews, van Driel, \& Gallagher (1998). 
Axes are heliocentric 
radial velocity, in \kms, versus flux density
in millijanskys.}

\figcaption{Rotation curve of UGC~7321 
derived from the VLA \HI\ aperture synthesis data of Rots,
Roberts, \& Goad (unpublished).
Axes are distance along the major axis, in arcseconds, versus heliocentric
radial velocity, in kilometers per second. The horizontal dashed line denotes
the systemic velocity measured from our
\nan\ \HI\ pencil beam measurements.
The arrows indicate the edge of the stellar disk in our WIYN $R$-band
observations, and the  crosses denote 
the limiting extent of the optical rotation curve derived  by
 Goad \& Roberts (1981). The close agreement between the location of
$V_{max}$ of the rotation curve and the global \HI\ profile horns in
Fig.~18 suggests are radial velocities are accurate to $\sim$10~\kms.}

\figcaption{Schematic drawing illustrating a thin disk inclined at
some angle $i$ to the line of sight.}

\figcaption{Blow-up of the central portion of the disk shown in
Fig.~20.}

\end{document}